\documentstyle[twocolumn,aps,prl,floats,epsf]{revtex}

\begin{document}

\draft

\wideabs{

\title{New Limit for
       the Family-Number Non-conserving Decay
       $\mu^{+} \rightarrow e^{+}\gamma$}

\author{
M.L.~Brooks,$^e$
Y.K.~Chen,$^c$
M.D.~Cooper,$^e$
P.S.~Cooper,$^b$
M.~Dzemidzic,$^c$
A.~Empl,$^c$
C.A.~Gagliardi,$^g$
G.E.~Hogan,$^e$
E.B.~Hughes,$^f$
E.V.~Hungerford~III,$^c$
C.C.H.~Jui,$^f$
J.E.~Knott,$^d$
D.D.~Koetke,$^h$
M.A.~Kroupa,$^e$
K.A.~Lan,$^c$
R.~Manweiler,$^h$
B.W.~Mayes~II,$^c$
R.E.~Mischke,$^e$
L.E.~Piilonen,$^j$
T.D.S.~Stanislaus,$^{e,h}$
K.M.~Stantz,$^d$
J.J.~Szymanski,$^{d,e}$
R.E.~Tribble,$^g$
X.L.~Tu,$^g$
L.A.~Van~Ausdeln,$^g$
W.H.~von~Witsch,$^c$
S.C.~Wright,$^a$
and K.O.H.~Ziock$^i$\\ \vspace*{9pt}
(MEGA Collaboration)\\ \vspace*{9pt}
}
\address{
$^a$University of Chicago, Chicago, IL 60637\\
$^b$Fermi National Accelerator Laboratory, Batavia, IL 60510\\
$^c$University of Houston, Houston, TX 77204\\
$^d$Indiana University, Bloomington, IN 46383\\
$^e$Los Alamos National Laboratory, Los Alamos, NM 87545\\
$^f$Stanford University, Stanford, CA 94305\\
$^g$Texas A \& M University, College Station, TX 77843\\
$^h$Valparaiso University, Valparaiso, IN 46383\\
$^i$University of Virginia, Charlottesville, VA 22901\\
$^j$Virginia Polytechnic Institute and State University, Blacksburg, VA 24061\\
}
\date{\today}

\maketitle

\begin{abstract}
An experiment has been performed to search for the
muon- and electron-number non-conserving decay
$\mu^{+} \rightarrow e^{+}\gamma$.  The upper
limit for the branching ratio is found to be 
$\Gamma(\mu^{+} \rightarrow e^{+} \gamma)$/
$\Gamma(\mu^{+} \rightarrow e^{+} \nu \overline{\nu})
\:<\:\:1.2\:\times\:10^{-11}$ with 90\%
confidence.
\end{abstract}

} 


	It is generally believed that the standard model
of electroweak interactions is a low-energy
approximation to a more fundamental theory.  Yet there is no clear
experimental evidence either to guide its extension
to additional physical processes or to predict the 
model parameters.
One of these model assumptions is lepton family-number conservation, which has been
empirically verified to high precision but is not a 
consequence of a known gauge theory.
Indeed many theoretical extensions to the standard model 
allow lepton-family-number violation within a range 
that can be tested by experiment~\cite{decay_list}.

      The predictions of the rate for a given family-number non-conserving process
vary among these extensions, and the most sensitive process depends on the model.
Many possibilities have been explored, and the present experimental limits for
a wide variety of processes have been tabulated in Ref.~\cite{cooper97}.
Of these, the 
rare muon decays have some of the lowest branching-ratio (BR) limits
because muons can be copiously produced and
have relatively long lifetimes.   The rare process,
$\mu^{+} \rightarrow e^{+} \gamma$, is the classic example of a 
reaction that would be allowed except for muon and electron number conservation;
the previous limit on the branching ratio is
BR($\mu^{+} \rightarrow e^{+} \gamma$) $<\:4.9\:\times\:10^{-11}$~\cite{bolton88}.
This decay is particularly sensitive to the standard
model extension that involves
supersymmetric particles~\cite{decay_list}.

	We report here a new limit for the BR of the decay
$\mu^{+} \rightarrow e^{+} \gamma$ from the analysis of data 
taken by the MEGA experiment at the Los Alamos Meson Physics 
Facility, LAMPF.  The dominant source of background in high-rate
$\mu^{+} \rightarrow e^{+} \gamma$
experiments is random coincidences between high-energy 
positrons from the primary decay process, 
$\mu^{+} \rightarrow e^{+} \nu \overline{\nu}$, and high-energy
photons from internal bremsstrahlung (IB), $\mu^{+} \rightarrow
e^{+} \gamma \nu \overline{\nu}$.  MEGA isolates the
$\mu^{+} \rightarrow e^{+} \gamma$ process from the background by
identifying the signature of the process: a 52.8-MeV photon
and a 52.8-MeV positron that are aligned back to back, in time coincidence,
and arise from a common origin.  Therefore, quality position, timing, 
and energy information are crucial.  In comparison to the
detector used to set the
previous limit~\cite{bolton88}, the MEGA detector sacrifices larger acceptance and
efficiency for better resolution, background rejection, and rate capability. 
It has been described in
several papers~\cite{photon_paper,trig_paper,ch_papers}
and will be discussed only briefly
below.

	Muons for the experiment are provided by a 
surface muon beam at the stopped muon channel at LAMPF.  
The muons, which are nearly 100\% polarized,
are brought to rest in a 76 $\mu$m Mylar foil, centered in
the 1.5-T magnetic field of a superconducting solenoid.
The angle between the muon beam and the normal to the target plane
is $82.8^{\circ}$ 
so that the stopping power
in the beam direction is increased, while the thickness of 
material presented to the decay positrons is minimized.  A sloped target
plane also extends the stopping distribution along the beam, enhancing the
sensitivity of the apparatus to the measurement of the decay position, 
which is the intersection of the outgoing photon 
and positron trajectories with the target foil.

	The positron and photon detectors are placed in the 
1.8-m diameter and 2-m axial length bore of the solenoid.  
Decay positrons from stopped muons
are analyzed by a set of high-rate, cylindrical multiwire-proportional
chambers (MWPC) surrounding 
the target.   They consist of
seven MWPCs arranged symmetrically outside of a larger MWPC,
coaxial with the central axis of the beam.
These MWPCs have a thickness of
$\:3\:\times\:10^{-4}$ radiation lengths,
minimizing energy loss while maintaining high acceptance and efficiency 
under the stopping rates of the experiment~\cite{ch_papers}.  The azimuthal location
of a passing charged particle is determined by anode wire readout.
The position of an event in the axial direction is obtained from
the signal induced on stereo strips scribed on the inner and outer
cathode foils of the MWPCs.  The positrons come to rest at either end of
the spectrometer in thick, high-$Z$ material
after passing through a barrel of 87 scintillators used for timing.
Outside these MWPCs, photons are detected in one
of three coaxial, cylindrical pair spectrometers~\cite{photon_paper}.
Each pair spectrometer consists 
of a scintillation barrel, two 250-$\mu$m Pb conversion foils sandwiching
an MWPC, and three layers of drift chambers, 
with the innermost having a delay-line readout to determine the axial 
position of a hit.

	The hardware trigger, consisting
of two stages of specially-constructed, high-speed
logic circuits, is fed signals from each of the three
photon spectrometers~\cite{trig_paper}.  Using pattern recognition programmed on
the basis of Monte Carlo (MC) simulations, the trigger requires
an electron-positron pair that can be potentially reconstructed
as arising from a photon of at least 37 MeV.
Since the instantaneous muon stopping rate in this experiment
is 250 MHz, with a macroscopic duty 
cycle of 6-7\%, the positron chambers and scintillators
have too many hits at any given time to be part of the
trigger.  Signals are digitized in FASTBUS with 6\% dead time
at the instantaneous trigger rate of 18 kHz.
Between each macropulse (120 Hz) of the accelerator, the data are read into one
of eight networked workstations, where an on-line algorithm
reduces the data rate for storage on magnetic tape to roughly 60 Hz.

	Each event is characterized by 5 kinematic parameters:
photon energy ($E_{\gamma}$), positron energy ($E_{e}$), relative time between
the positron and photon ($t_{e \gamma}$) at the muon decay point,
opening angle ($\theta_{e \gamma}$), and photon traceback angle ($\Delta \theta_{z}$).
These properties, in conjunction with the detector response, determine
the likelihood that a signal is detected.
The determination of the detector acceptance and response
functions relies on a MC simulation to
extrapolate from experimental input to the kinematic region of
the $\mu^{+} \rightarrow e^{+} \gamma$ signal.  To verify
the MC calculation, a number of
auxiliary measurements are performed.
The two most important are the $\pi^{-}_{stopped}
p \rightarrow \pi^{0} n \rightarrow \gamma \gamma n$ process and
the prompt $e- \gamma$ coincidence signal from the IB decay.

\begin{figure}
  \begin{center}
    \mbox{\epsffile{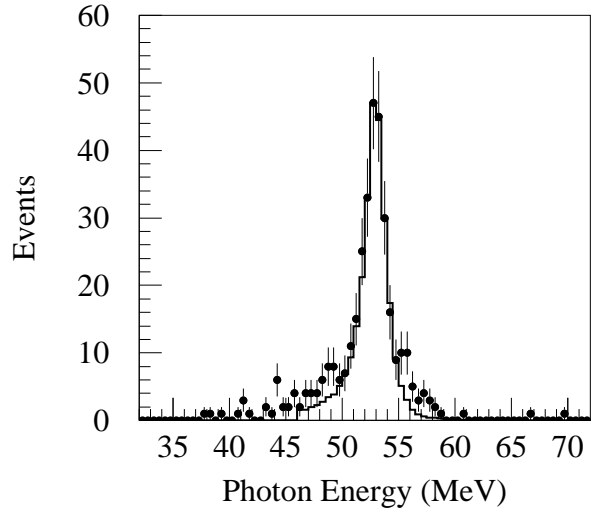}}
  \end{center}

  \caption{The $E_{\gamma}$ spectrum from photons converting in the outer
  layer of lead.  The data points are produced by stopping pions in $CH_{2}$ via
  the reaction $\pi^{-} p \rightarrow \pi^{0} n \rightarrow \gamma \gamma n$,
  scaled from 54.9 to 52.8 MeV.  The curve is the response function generated
  from the MC and used
  in the analysis of the $\mu^{+} \rightarrow e^{+} \gamma$ data.}

  \label{fig:pions}
\end{figure}

	Pion capture at rest on hydrogen produces photons with energies between
54.9 and 83.0 MeV.  Under the condition that the two photons
have a minimum opening angle of 173.5$^{\circ}$, these photons are restricted
to have energies close to 54.9 and 83.0 MeV and a spread much smaller
than the detector response.  Figure~\ref{fig:pions} shows
the experimental line shape for the 54.9 MeV photon for conversions in the
outer Pb foils of the three pair spectrometers, scaled to 52.8 MeV.
The curve is the response function generated from the MC that is used in the analysis
of the $\mu^{+} \rightarrow e^{+} \gamma$ data.
We attribute differences in the low-energy
tail to charge exchange of in-flight pions from carbon in the $CH_{2}$ target and
discrepancies in the high-energy tail to contributions from other
opening angles due to reconstruction problems for the high-energy photon.
The measured and simulated line shapes agree better for conversions in
inner Pb foils, which have worse resolution.
The energy resolutions are 3.3\% and 5.7\% (FWHM) at 52.8 MeV
for conversions in the outer and the inner Pb layers, respectively.
The $\pi^{\circ}$ decays also provide the time response between
the two photons, which is
reasonably characterized by a Gaussian with a $\sigma$ = 0.57 ns for each photon.

\begin{figure}
  \begin{center}
    \mbox{\epsffile{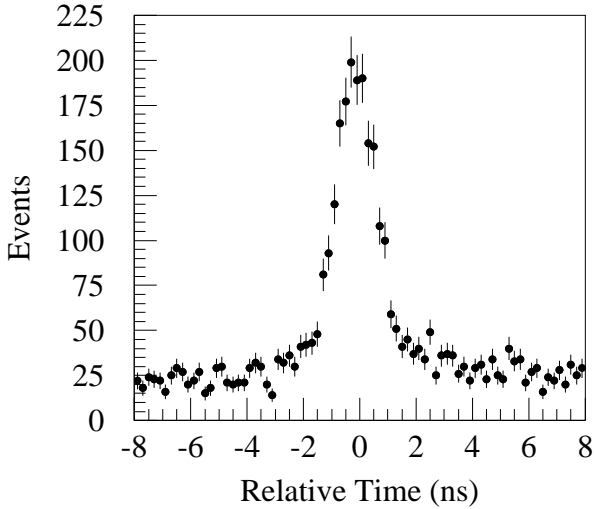}}
  \end{center}

  \caption{Values for $t_{e \gamma}$ from the process
  $\mu^{+} \rightarrow e^{+} \gamma \nu \overline{\nu}$ under the conditions
  of reduced rate and magnetic field.}

  \label{fig:ib}
\end{figure}

     Observation of the IB process demonstrates that the apparatus can detect
coincident $e- \gamma$ events.  At
nominal beam intensity, this process is completely engulfed by random
coincidences.  Figure~\ref{fig:ib} shows the spectrum for $t_{e \gamma}$, with the
beam intensity reduced by a factor of 60, the magnetic field lowered by 25\%,
and the $\mu^{+} \rightarrow e^{+} \gamma$
online filter suppressed.  The peak
shown is for all energies of the detected decay products.  The area of the
peak is very sensitive to the exact acceptances of the detector at its
thresholds and can be calculated by MC simulation to within a factor of two.  If the data
and the simulation are restricted to $E_{\gamma} >$ 46 MeV,
$E_{e} >$40 MeV, and $\theta_{e \gamma} >120^{\circ}$, the branching
ratio is reproduced within uncertainties.  The
shape of the peak can be characterized by a Gaussian with a
$\sigma$ = 0.77 ns.  The dominant contributor is
the photon timing, as measured in the stopping-pion experiment, which
must be scaled down from about 70 to 40 MeV for the comparison.  At 52.8 MeV, the
MC simulation indicates the photon-positron resolution is $\sigma$ = 0.68 ns.

         In the IB and $\mu^{+} \rightarrow e^{+} \gamma$ processes,
the origin of the photon is defined
to be the intersection of the positron with the target.
The photon traceback angle, $\Delta \theta_{z}$, specifies
the difference between the polar angles of the photon as determined
from the line connecting the decay point to the photon conversion point
and from the reconstructed $e^{+}-e^{-}$ pair.
The resolution of $\Delta \theta_{z}$ is dominated
by multiple scattering of the pair in the Pb converters.
The observed response
for inner and outer conversion layers for the IB process is in
excellent agreement with the MC simulation.
The traceback resolutions appropriate for the $\mu^{+} \rightarrow e^{+} \gamma$
analysis are $\sigma$ = 0.067 and 0.116 rad
for conversions in the outer and the inner Pb layers, respectively.

\begin{figure}
  \begin{center}
    \mbox{\epsffile{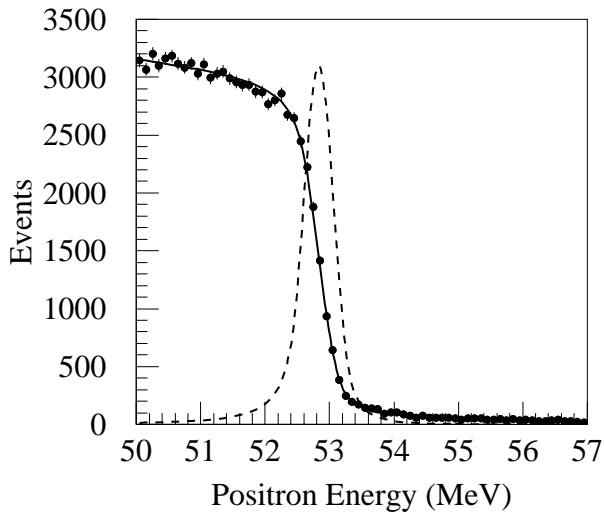}}
  \end{center}

  \caption{The $E_{e}$ spectrum from
  $\mu^{+} \rightarrow e^{+} \nu \overline{\nu}$ extracted from full rate
  data for the middle topology group.
  The solid curve is the fit used to extract the line shape (dashed curve).}

  \label{fig:electron}
\end{figure}

         The resolution of $E_{e}$ is determined by 
the slope of the high-energy cut-off edge in the spectrum of the decay,
$\mu^{+} \rightarrow e^{+} \nu \overline{\nu}$.  It
depends on the ``topology'' of the track, which is determined by
the number of loops these particles make in 
the magnetic field between the target and scintillator
and the number of chambers they traverse.
The $E_{e}$ spectrum is shown in Fig.~\ref{fig:electron}
for one of three topology groups.
The MC
line shape is characterized near the centroid
by a Gaussian and in the tails by different powers of the deviation
from the central energy.  To extract the response function
from the data, this line shape is convoluted
with the spectrum from normal muon decay, modified by detector acceptance
and unphysical ``ghost'' tracks. Ghost tracks are a high-rate phenomenon and
are reconstructions made from the fragments of several physical
tracks.  They are the source of events well above the kinematic
limit for the positron energy.
The solid curve in Fig.~\ref{fig:electron} is the fit, and the dashed curve
is the corresponding line shape.  The central Gaussians of the three
topology groups have $\sigma$ =
0.21, 0.23, and 0.36 MeV.

	There is no way to measure the response function for $\theta_{e \gamma}$.
The MC simulation is relied upon to produce this
distribution and gives 
the FWHM for cos($\theta_{e \gamma}$) as
$1.21\:\times\:10^{-4}$ at $180^{\circ}$.
Given helical tracks, knowing the
location of the target is critical to obtaining the correct absolute value of
$\theta_{e \gamma}$, and the mechanical survey provides the most accurate
measurement for the analysis.

        The data for this experiment have been taken in three calendar years, 1993-95.
The full
data set is based on $\:1.2\:\times\:10^{14}$ muon stops collected over $\:8\:\times\:10^{6}$ s of
live time and results in $\:4.5\:\times\:10^{8}$ events on magnetic tape.
These events are passed through a set of computer
programs that reconstruct as many as the pattern recognition
algorithms can interpret.  The programs include physical effects such
as mean energy loss in matter and non-uniformities in the magnetic
field.  Events are required to satisfy separate $\chi_{\nu}^2$ cuts
on the positron and photon fits and loose cuts on the signal kinematics
($E_{e}\:>\:$50 MeV, $E_{\gamma}\:>\:$46 MeV, $|t_{e \gamma}|\:<\:$4 ns,
cos($\theta_{e \gamma}$)$\:<\:$-0.9962, and $|\Delta \theta_{z}|\:<\:$0.5 rad).
Events in which the positron momentum vector at the decay point appears
to lie within 5$^{\circ}$ of the plane of the target are discarded.  After
roughly one year of computing on a farm of UNIX workstations, the data set
has been reduced to 3971 events that are fully reconstructed and
of continuing interest.  This sample is large enough to allow
a study of the background.
To remove incorrectly reconstructed events,
the images of the photon showers in the pair
spectrometers are manually scanned.  The efficiency
for keeping real photons is monitored by mixing about 500 52.8-MeV MC events into
the sample in a non-identifiable
way and finding that 91\% of the MC events pass, whereas only 73\% of
the data events are selected.

      The acceptance of the apparatus is obtained by simulating
$\:1.2\:\times\:10^{7}$ unpolarized $\mu^{+} \rightarrow e^{+} \gamma$ decays
and finding that $5.2\:\times\:10^{4}$ events survive processing by the same
codes used for the data analysis.  Thus the probability that a $\mu^{+} \rightarrow
e^{+} \gamma$ decay would be detected is $\:4.3\:\times\:10^{-3}$.  This value
is reduced by 20\% to account for inadequacies in the MC simulation that
over estimate the acceptance.  The shortcomings
primarily involve inter-channel cross talk and are estimated by comparing the
images of many data and MC events.  The acceptance is further reduced by 9\%
for the inefficiency of
manual scanning. The total number of muon stops is determined by calibrating 
the rates in the positron scintillators to a known muon flux.  After
correcting for dead time, the single event sensitivity for the experiment is
$\:2.3\:\times\:10^{-12}\:$=$\:1/N_{\mu}$, where $N_{\mu}$ is the number of
useful stopped muons.

     The determination of the number of
$\mu^{+} \rightarrow e^{+} \gamma$ events in the sample is evaluated using
the likelihood method described in the analysis of previous
experiments~\cite{kinnison82}.
The formula for the normalized likelihood is
\begin{displaymath}
{\cal L}(N_{e \gamma}) = 
\prod_{i=1}^{N} \left( \frac{N_{e \gamma}}{N} \left( \frac{P}{R}-1 \right)
+\frac{N_{IB}}{N} \left( \frac{Q}{R}-1 \right)
+1 \right),
\end{displaymath}
where $N=3971$, $N_{e \gamma}$ is the number of
signal events, $N_{IB}$ is the number of IB events, and $P$, $Q$, and $R$ are the
probability density functions (PDF) for signal, IB, and randoms of each
of the five parameters describing
the event.
The PDFs $P$ and $R$ are the products of statistically
independent PDFs for the five parameters, each normalized to
unit probability over the full range of the variable for
the sample.  The signal distributions are taken
from MC distributions as described. The background PDFs
are extracted from the spectral shapes of a much larger sample of events,
where the constraints on the other statistically independent parameters
remain very loose.  Here $Q$ is taken from MC simulation of the IB
and has correlations amongst the variables.
The events fall into the following categories:  positron topology, photon
conversion plane, target intersection angle, and data taking period.
As a result, PDFs are extracted for each class of events
and applied according to the classification of individual events.

     The likelihood function evaluates the statistical separation between signal,
IB, and background.
To observe the impact of quality constraints in the
pattern recognition, they have been relaxed
to produce a sample three-times larger.
One event emerges with a large value of $P/R$ that is significantly
separated from the distribution.  
However, this event has a large positron $\chi_{\nu}^2$,
indicative of a ghost track.  The adopted constraints produce a sample
with considerably less background.
The peak of the likelihood function is at $N_{e \gamma}$=0 and $N_{IB}$=30$\pm$8$\pm$15.
The systematic error assigned to $N_{IB}$ is due to the uncertainty in the shape
of the background time spectrum when the events are filtered by the online program.
The expected number of IB events is 36$\pm$3$\pm$10, where the systematic error is due
to finite resolution effects across the cut boundaries.
The 90\% confidence limit
is the value for $N_{e \gamma}$ where 90\% of the area of the
likelihood curve lies below $N_{e \gamma}$ and $N_{IB}$ is maximal.  This value is
$N_{e \gamma}<5.1$.  Therefore, the limit on the branching ratio is
\begin{displaymath}
\left. \frac{\displaystyle \Gamma(\mu \rightarrow e \gamma)}
{\displaystyle \Gamma(\mu \rightarrow e \nu \overline{\nu})} \right. \leq 
\frac{\displaystyle 5.1}{\displaystyle N_{\mu}} = \:1.2\:\times\:10^{-11}\:
(90\%\:CL).
\end{displaymath}
In comparison to the previous experimental limit~\cite{bolton88}, this
result represents a factor of 4.1 improvement.  The previous experiment
would have had 58 background events at the same sensitivity instead
of the 2 found here.  This improvement
further constrains attempts to build extensions to the standard
model~\cite{decay_list}.

	We are grateful for the support received by LAMPF staff members 
and in particular, P. Barnes, G. Garvey, L. Rosen, and D. H. White.  We wish to gratefully
acknowledge the contributions to the construction and operation of this
experiment from former collaborators, the engineering and technical staffs,
and undergraduate students
at the participating institutions.
The experiment is supported in part by the US 
Department of Energy and the National Science Foundation.


\begin{references}

\vspace*{-24pt}
\bibitem{decay_list}R. Barbieri, L. Hall and A. Strumia, 
Nucl. Phys. {\bf B455}, 219 (1995); N. Arkani-Hamed, H-C. Cheng, and 
L. Hall, Phys Rev {\bf D53}, 413 (1996); T. Kosmas, G. Leontaris 
and J. Vergados, Prog. Part. and Nucl. Phys. {\bf 33}, 397 (1994) and
included references. 
\bibitem{cooper97}M. Cooper, {\it et. al.}, Proc. of the 6$^{th}$ Conf. on the
Intersections between Particle and Nuclear Physics, AIP, T. W. Donnelly, ed,
Big Sky, MT 34 (1997).
\bibitem{bolton88}R. Bolton, {\it et. al.}, Phys. Rev. {\bf D38}, 2077 (1988).
\bibitem{photon_paper}M. Barakat, {\it et. al.}, Nucl. Inst. and 
Meth., {\bf A349}, 118 (1994).
\bibitem{trig_paper}Y. Chen, {\it et. al.}, Nucl. Inst. and 
Meth., {\bf A372}, 195 (1996).
\bibitem{ch_papers}V. Armijo, {\it et. al.}, Nucl. Inst. and Meth.,
{\bf A303}, 298 (1991); T. D. S. Stanislaus, {\it et. al.}, Nucl. Inst. and 
Meth., {\bf A323}, 198 (1992); M. D. Cooper, {\it et. al.}, Nucl. Inst. and 
Meth., {\bf A417}, 24 (1998).
\bibitem{kinnison82}W. W. Kinnison, {\it et. al.}, Phys. Rev., {\bf D25},
2846 (1982).
\end{references}
\end{document}